\documentstyle[12pt]{article}
\begin{document}
\setcounter{page}{0}
\thispagestyle{empty}
\baselineskip22pt plus3pt
\begin{center}
{\Large{\bf Maximal Lyapunov exponent at Crises}}\\
\vspace{1cm}
{Vishal Mehra and Ramakrishna Ramaswamy\\}
{\it School of Physical Sciences\\
Jawaharlal Nehru University, New Delhi 110 067, INDIA\\}
\vspace{1cm}
\today
\end{center}
\begin{abstract}

We study the variation of Lyapunov exponents of simple dynamical
systems near attractor-widening and attractor-merging crises.
The largest Lyapunov exponent has universal behaviour,  showing
abrupt variation as a function of the control parameter as the
system passes through the crisis point, either in the value
itself,  in the case of the attractor-widening crisis, or in
the slope,  for attractor merging crises. The distribution of
local Lyapunov exponents is very different for the two cases:
the fluctuations remain constant through
a merging crisis, but there is a dramatic increase in the
fluctuations at a widening crisis.

\end{abstract}
\vspace{1cm}
\newpage
\section*{I. INTRODUCTION}

In this paper, we study the behaviour of the Lyapunov exponent
in systems where there are abrupt changes in the dynamics
as a parameter is varied. Our interest is in exploring
the typical dependence of the maximal Lyapunov exponent (MLE) on
the control parameter so as to elucidate the signature of a
transition in the nature of the dynamics.

In the context of dynamical systems, abrupt changes in the phase
space most commonly occur at the so-called {\it crises}
\cite{GORY87}, which are caused by the collision of a chaotic
attractor with the stable manifold of an unstable periodic
orbit. The three major types of crises are distinguished by the
nature of discontinuous change they induce in the chaotic
attractor. At a {\it boundary crisis,} the chaotic attractor is
suddenly destroyed and replaced by a chaotic transient as the
parameter passes through its critical value. This occurs when
the attractor collides with the stable manifold of an unstable
periodic orbit that lies on its basin boundary.  At an {\it
interior crisis,} a sudden increase or decrease in the size of
the attractor occurs when the stable manifold of an unstable
periodic orbit lying within the basin of attraction of the
chaotic attractor collides with it.  At an {\it
attractor-merging crisis } two or more chaotic attractors
simultaneously collide with the stable manifold of an unstable
periodic orbit lying on their common basin boundary which
results in the merging of the attractors.

The qualitative change in the dynamics at a crisis is
reflected in the Lyapunov exponents.  The case of a
boundary crisis is not very interesting since the Lyapunov exponent
is either zero (if the transient leads to a periodic attractor) or
takes a value characteristic of the chaotic attractor onto which the
trajectory ultimately lands. The variation of the MLE at typical
interior and merging crises is more dramatic, and in this paper we
study these phenomena in a variety of simple model systems
\cite{M76,KSE85,H76,GH83}. In all the crises, there
is a similarity in the
dependence of the Lyapunov exponent on the control parameter. Our
major observation is that MLE has a characteristic behaviour which
is, however, distinct for the attractor-widening and
attractor-merging cases.  For interior crises which terminate a
periodic window the dependence of MLE on the control parameter is
sigmoidal, with a {\it large} increase in fluctuations subsequent to
the crisis. This abrupt increase in the MLE at interior crises has
been observed before \cite{LK81,RH84,GW86,PL88,ST94} in some
studies of $1-d$ and $2-d$ maps and flows.  In contradistinction the
MLE only has a ``knee'' at attractor-merging crises: after the
crisis, the rate of change of the Lyapunov exponent decreases
significantly. Again, in contrast to the attractor-widening case,
there is no attendant increase in the fluctuations of the local
Lyapunov exponents subsequent to the crisis.

In the next section we describe the phenomenology of the
behaviour of the MLE at crises in simple  maps and other low
dimensional dynamical systems. Our results have relevance to
studies of systems at a phase transitions, especially as a
number of recent simulations of realistic systems have looked at
the Lyapunov exponent, K-entropy and related quantities as a
function of temperature or other control parameters
\cite{BLB88,BC87,SRC95}.  These
considerations are discussed in relation to the present work in
the concluding Section~III.

\section*{II. The Lyapunov exponent at Crisis}

The Lyapunov exponent, which is used to characterise the degree of
chaoticity of a dynamical system gives the average rate of
exponential divergence of two nearby trajectories \cite{LL92}. In a $
n-\/$dimensional dynamical system there are $n$ Lyapunov exponents
and the system is chaotic if at least one of them is positive while
for regular dynamics all Lyapunov exponents are zero or negative.  We
focus on the largest of these, which is most simply defined as
\begin{equation}\label{define}
\lambda_m=\lim_{t\rightarrow\infty}\lambda_m(t)
=\lim_{t\rightarrow\infty}\frac{1}{t}
\lim_{d(0)\rightarrow 0}\log\frac{d(t)}{d(0)}
\end{equation}\noindent
where $d(0)$ is the initial separation between two trajectories, and
$d(t)$ is their separation after time $t$. A number of methods have been
proposed in the literature to compute one or more of the Lyapunov
exponents\cite{BGS76,SN79,BLB88}. Here we use the
tangent space method\cite{BGS76} which is sufficient since we are
interested primarily in the largest Lyapunov exponent.

To study transient objects like repellers or semi-attractors one can also
analyse the  finite-time exponents $\lambda_m(t)$ which are also defined
in Eq.~\ref{define}. The instability fluctuations
on an attractor can also be studied by  dividing a long ergodic trajectory in
segments of size $t$ and calculating the Lyapunov exponent
$\lambda_m(t)$ for each of these.
The probability density $P(\lambda_m(t))$ of the distribution of
local Lyapunov exponents has the scaling form, for $t\rightarrow\infty$,
\begin{equation}
P(\lambda_m(t))\sim\exp(-t\psi(\lambda_m(t))),
\end{equation}\noindent
where $\psi(\lambda_m(t))$ is a concave function with its
minimum equal to zero at $\lambda_m=\lambda_m(\infty)$.
\cite{MMYMHH,GBA88}. In a
highly mixing system, the time correlations of $\lambda_m(t)$ can be
ignored and then by the central limit theorem $P(\lambda_m(t))$ is a
gaussian and $\psi(\lambda_m)$ is parabolic. However at crises points
the gaussian distribution breaks down and $\psi(\lambda_m(t))$
develops a cusp at its minimum \cite{MMYMHH,THHMM}.

\subsection*{A. Interior or widening crises }
We first consider the logistic map
\begin{equation}
x_{n+1}=rx_n(1-x_n)
\end{equation}\noindent
It is well-known\cite{M76} that as the parameter $r$ is
increased the logistic map undergoes period-doubling cascade
terminating at the accumulation point $r_{\infty}
\simeq~3.5699\ldots$~. Beyond that the dynamics is mainly
chaotic, punctuated at various intervals by periodic
windows of arbitrarily high period. An odd period-$n$ window is
created at a saddle-node bifurcation together with an unstable
period-$n$ orbit. As $r$ is increased there is a sequence of
period-doubling bifurcations creating periodic attractors of
period $2n,~2^2n,~2^3n,~\ldots$~. Beyond the accumulation point of
the period-doubling bifurcations the attractor is made up of $n$
distinct pieces. The trajectory hops among these pieces in a
regular manner but the distribution of points within each piece
is random on the so-called semi-periodic attractor\cite{KST93}.
At the right end of the window there is an interior crisis when
each piece of the semi-periodic attractor meets a point of the
unstable period-$n$ orbit which was created in the saddle-node
bifurcation, leading to an abrupt increase in the accessible
phase space volume \cite{GORY87}.

Shown in Fig. 1 is the variation of $\lambda_m$ with $r-r_c^n$ near
the $n=3,5$ and 7-band crises, which occur at parameter $r=r_c^n$
respectively.  We observe that for all interior crises the MLE {\it vs.~}
$(r-r_c^n)$ curve is sigmoidal. The standard deviation in the
local Lyapunov exponent (calculated from $N= 10^4$ with 50 different
initial conditions) increases dramatically at the crisis.  It is easy
to see why the fluctuation in the local exponents should increase
abruptly at the interior crises: the attractor gains large volume
which may have entirely different stability properties
\cite{ST94,KST93}. The MLE increases at the crisis
because the attractor engulfs the coexisting repeller. This repeller
is the remnant of the chaotic attractor which had ceased to exist at
the saddle-node bifurcation.  Computation of finite-time Lyapunov
exponent near crisis shows that the repeller has larger finite-time
Lyapunov exponent than the semi-periodic attractor\cite{PL88}.  The
spectrum of local Lyapunov exponents of the post-critical attractor
just before and after the 3-band crisis is shown in Fig. 2(a).
The linear segments indicating non-hyperbolicity at the crisis are
present on both sides of $r_c$. After the crisis a
kink at large $\lambda$ is visible which corresponds to the
distribution on  the repeller \cite{MMYMHH}.

Pompe and Leven \cite{PL88} argued that the increase in MLE is
proportional to the probability density on the repeller, and model it
by the power-law
\begin{equation}
P_R\sim(r-r_c)^{\mu}
\end{equation}\noindent
We confirm that the increase in MLE is proportional to the probability density
on the repeller. A power law fit to MLE data gives the exponent
$\mu=0.51\pm 0.04$ for 3-band crisis and $\mu=0.52\pm0.04$ for 5 and
7 band crises.  Indeed, Grebogi {\it et al.} \cite{GOY83} obtained an
approximate scaling near $r_c^3$
\begin{equation}
P_R\sim(r-r_c^3)^{1/2}g(\ln(r-r_c^3)),
\end{equation}\noindent
where $g$ is a periodic function. It therefore appears that this scaling
relation is valid at all other band-crises as well.

Other $1-d$ maps also show the same phenomenology.  We have studied a
map originally introduced by Kariotis, Suhl and Eckmann\cite{KSE85} in
order to mimic the dynamical behaviour of intramolecular processes
and isomerization. This map is given as
\begin{equation}\label{KSE85}
x_{n+1}=rx_n(\omega^3-2\omega x_n^2+x_n^4),
\end{equation}\noindent
We fix $\omega=0.8$ and consider $r$ as the control parameter.  The
above map shows both attractor-merging (discussed below) and
attractor-widening crises. At the 5-band crisis at $r_c^5 =
5.2505109\ldots$.,  as shown in Fig. 1,
clearly shows that the behaviour of the Lyapunov exponent is
essentially identical to that observed for the logistic map.

Higher dimensional systems also show the same behaviour: for example,
the well known H\'enon map,
\begin{eqnarray}
x_{n+1}&=&y_n+1-rx_n^2 \nonumber\\ y_{n+1}&=&bx_n
\end{eqnarray}\noindent
which has a well-characterized, complex structure of bifurcations and
crises\cite{LL92}. Fixing $b$=0.3 and varying $r$, the 7-band crisis
occurs at $r_c^7$=1.2716856.  Again (cf. Fig. 1) it is seen that the
maximal Lyapunov exponent shows the by now familiar characteristic
sigmoidal behaviour as the function of $(r-r_c^7)$.

\subsection*{B. Attractor-merging crises}
Attractor-merging crises typically occur because of some symmetry in
the underlying dynamics, for example in the logistic map, beyond the
accumulation point of the period-doubling cascade $r_{\infty}$ there
is a successive merging of chaotic bands.  Thus, for $r<r_m$ the
chaotic attractor consists of $2^m$ chaotic bands.  If we take a
point in any one of these $2^m$ bands the trajectory will come to
that band after $2^m$ iterations. So that the band can be considered
as an attractor for $2^m$-times iterated map.  So the band-merging
phenomenon can be regarded as an attractor-merging crisis\cite{GORY87}
for the $2^m$-times iterated map.

We show results for the $m = 1$ merging at $r_m = 3.678486 \ldots$
and the $m~=~2$
merging at $r_m = 3.5925663 \ldots$ in Fig. 3.  For band-merging
crises the MLE {\it vs.~}  $(r-r_m)$ curve has a sharp {\it knee}
precisely at $r =
r_m$, {\it i.e.}, the derivative of MLE is discontinuous at $r_c$.
However the knee angle is not same for all band-mergings.  The local
Lyapunov exponents are more uniform here, and we do not see any
significant change in the fluctuation properties after the crisis.
This is not difficult to understand as the co-merging attractors are
symmetry-related.  Similar behaviour was observed for attractor-mergings
in the other maps studied \cite{KSE85,H76} (Fig.~3).
 The spectrum of local Lyapunov exponents near the
attractor-merging crisis (Fig 2(b)) shows the linear segments indicative of the
non-hyperbolicity at the crisis \cite{THHMM}.

We also consider a merging-crisis in the forced Duffing equation \cite{GORY87}
\begin{equation}\label{duff}
d^2x/dt^2+\nu dx/dt+\alpha x^3-\beta x=r\sin \omega t
\end{equation}\noindent
We take $\nu=1, \alpha=100, \beta=10, \omega=3.5$ and study Eq.
(\ref{duff}) near its crisis value $r_m\approx0.853$. Below $r_m$ there are
two chaotic attractors one confined to the well in $x>0$ and another
confined to the well in $x<0$. These attractors merge at $r=r_m$,
where the MLE near $r=r_m$ again shows a knee (Fig. 3).

\section*{III. Conclusions}

In this work we have studied the variation of the largest Lyapunov
exponent MLE near interior and attractor-merging crisis for some
simple and well-known systems.

We observe that around an interior crisis, the MLE {\it vs.~} the control
parameter curve has a sigmoidal dependence, with the fluctuations
increasing dramatically at the crisis. For the 3-band crisis in the
logistic map (Fig. 1), for example, the average fluctuations just before the
crisis is $4.2\times10^{-4}$, while after the crisis it is $1.2\times10^{-3}$.
On the other hand, a knee-shaped curve is observed for attractor-merging crises
 with no increase in fluctuations beyond crisis.

In  recent work  Fan and Chay\cite{FC95} have studied the Lyapunov
exponents of Rose-Hindmarsh system (consisting of three coupled
differential equations), and report that Lyapunov exponents are not
good indicators of an interior crisis. They prefer the use of
topological entropy which showed an abrupt increase at the crisis.
However, in contrast to the present systems where the interior crises
terminate a periodic window the interior crisis they studied was
caused by collision of two period-adding bifurcation processes
travelling in opposite directions in the parameter space. This may be
one reason why they did not observe increase in the MLE at the
crisis.

The observations made above are of relevance to recent simulation
studies of systems undergoing a change in bulk phase.  In recent
years it has become possible to study the detailed dynamics of
mesoscopic systems undergoing phase change, and a number of studies
\cite{BLB88,BC87,SRC95} have therefore focussed on the relation between
phase transitions and the Lyapunov exponents that characterise the
dynamics.  For example, in a study of a large number of coupled
planar rotors, Butera and Caravati\cite{BC87} found a discontinuity
in the slope of the maximal Lyapunov exponent (MLE) at the precise
temperature of the Kosterlitz-Thouless transition.  More recently, it
has been seen in molecular dynamics simulations of small
Lennard-Jones clusters\cite{SRC95} that the largest Lyapunov exponent
also increases dramatically as the system makes a transition from a
solidlike to a liquidlike phase. This abrupt change characterizes the
phase transition in a manner exactly analogous to the Lindemann
criterion, and indeed offers an alternative connection between the
phase space dynamics and the phase change.

This abrupt increase in the MLE corresponds to an increase in the
available phase space volume and consequently in the local rate of
divergence of trajectories\cite{BGS76}. Berry and
co-workers\cite{BLB88} have looked at a variety of dynamical
indicators, including the Kolmogorov-Sinai (KS) entropy, {\it i.e.},
the sum of all positive Lyapunov exponents.  This quantity increases
smoothly with temperature or energy as the phase changes although
information can be obtained regarding underlying potential-energy
surface\cite{BLB88}.

The present study further underscores the utility of the MLE as an
indicator of phase transformation since the band-crises between different
chaotic phases are, in a sense, dynamical system analogues of phase
transitions. Prior to the crisis there is long-time correlated noisy
periodicity while after the crisis, dynamics lacks long-time correlation
since motion is chaotic within a single-band attractor.

\section*{Acknowledgements:} This work was supported by grant
SPS/MO-5/92 from the Department of Science and Technology, and the CSIR,
India. We also thank C. Chakravarty for comments on the manuscript.

\newpage
\section*{Figure Captions}
\begin{enumerate}

\item[Fig. 1]
Variation of the maximal Lyapunov exponent , $\lambda_m$, vs. $(r-r_c)$
around interior crises. The error bars show the magnitude of the
fluctuation in the local Lyapunov exponent. Beyond the crisis, solid line
through the data is the power-law $r-r_c)^\mu$, with $\mu=0.5$ for a), b)
and c), and $\mu=0.44$ for d) and 0.37 for e). 3-band crisis  in logistic
map at $r_c^3=3.8568007$. (b) 5-band crisis in the logistic map at
$r_c^5=3.7447104$.  (c) 7-band crisis in the logistic map at
$r_c^7=3.70279404 $. (d) 5-band crisis in the Kariotis-Suhl-Eckmann map at
$r_c^5=5.2505109$. (e) 7-band crisis in the henon map at
$r_c^7=1.2716856$, $b$ is fixed at 0.3.

\item[Fig. 2]
Spectrum of local Lyapunov exponents, $\psi(\lambda(n))$, just
before and after a crisis, for logistic map (a) the 3-band crisis- $r =
r_c^3\pm\delta r$ . (b)the $m=2$ band-merging-- $r = r_m\pm\delta r$.
Here $n=60$ and the number of iterations of the map is $3\times 10^6$.
$\delta r=1\times 10^{-6}$.  Crosses refer to data before crisis and
circles to data after crisis.

\item[Fig. 3]
 Variation of the maximal Lyapunov exponent, $\lambda_m$, vs.
$(r-r_c)$ around attractor-merging crises. As in Fig. 1 the error bars
show the magnitude of the fluctuation in the local Lyapunov exponent. The
straight lines through the data are least-square fit constrained to pass
through the critical point.
(a) $m=2$ band-merging in the logistic map at $r=3.678486$. (b) $m=1$
band-merging in the logistic map at $r=3.59256296 $. (c) Attractor-merging
in Kariotis-Suhl-Eckmann map at $r=5.74009035 $. (d) H\`enon map-- merging
crisis at $r=1.084404$, $b$ is again fixed at 0.3.  (e) Well-merging in
the forced Duffing equation at $r\approx 0.853$.
\end{enumerate}
\end{document}